\documentstyle[aps,psfig]{revtex} 
\tightenlines 
\newlength{\figwidth} 
\title{Exact and semiclassical Husimi distributions of Quantum Map 
Eigenstates} 
\author{M. Saraceno$^1$ and A. G. Monastra$^2$} 
\address{$^1$Departamento de F\'{\i}sica, Comisi\'on Nacional de Energ\'{\i}a
At\'omica (CNEA), \\ Av. del Libertador 8250, 1429 Buenos Aires, Argentina
\\ $^2$Laboratoire de Physique Th\'eorique et Mod\`eles Statistiques,
B\^at. 100, \\ 91405 Orsay Cedex, France} 
 
\begin{document} 
\maketitle {\begin{abstract} The projector onto single quantum map eigenstates
is written only in terms of powers of the evolution operator, up to half the
Heisenberg time, and its traces. These powers are semiclassically
approximated, by a complex generating function, giving the Husimi distribution
of the eigenstates. The results are tested on the Cat and Baker maps.  
\end{abstract}}

\section{Introduction} 
The relation between the classical invariant structures of a dynamical 
system and its 
quantum counterparts, the eigenenergies and eigenstates, is not 
completely understood and 
is an intense subject of investigation. When the system is integrable 
the eigenenergies 
are determined by the well-known Bohr-Sommerfeld or EBK quantization 
rules and the 
eigenfunctions have a very intuitive phase space structure: they are 
essentially 
localized on the quantized tori and they present characteristic quantum 
features 
--interference fringes for the Wigner function \cite{berry1} or zeros 
for the Husimi 
function \cite{voros}-- away from them. No such clear picture is 
available for the 
eigenstates of chaotic systems and much work
\cite{heller,bogo1,vergini,nonnenmacher} has gone 
into trying to disentangle the purely statistical aspects of the 
eigenfunctions from the 
specific dynamical features such as scars and localization. 
 
Although a lot of recent work deals with the study of the statistical 
aspects of chaotic 
wavefunctions \cite{mirlin} we concentrate here on the 
purely dynamical 
aspects and explore how well the semiclassical formulae can reproduce 
the details of 
individual eigenfunctions. 
 
With regards to spectral properties the theoretical tool for this 
analysis is the 
Gutzwiller trace formula \cite{gutzw} relating the spectral density to 
the periodic 
orbits. As is well known, this relationship is fraught with 
uncertainties both due to 
the approximations involved, as to the fact that it diverges where it is 
supposed to 
locate the single eigenenergies and to the sheer complexity of adding 
the contributions 
of an exponentially increasing number of periodic orbits. Some of these 
difficulties can 
be overcome by resummation techniques \cite{bogo-keat} that limit the 
exponential 
increase by partially restoring the unitarity of the propagator, lost in 
the 
semiclassical approximation. 
 
In the present work we extend these resummation methods to the 
calculation of 
eigenfunctions of chaotic maps. We utilize Fredholm theory to construct 
a projector on 
single eigenstates whose expression is directly amenable to 
semiclassical 
approximations. These approximations share the same advantages - and the
drawbacks- of the resummation techniques available for spectral properties.
 
In section 2 we construct explicitly the projector on single 
eigenstates. This is 
achieved exactly in terms of a finite sum on powers of the quantum map 
propagator and 
its traces. The cut-off time is half the Heisenberg time due to the 
explicit imposition 
of unitarity. In section 3 these formulae are implemented 
semiclassically by a) 
approximating the traces in terms of actions and instabilities of 
periodic orbits in the 
usual way, and b) by computing the propagator in the coherent state 
representation in 
terms of specially devised complex generating functions. 
 
The final result is a semiclassical expression for the Husimi function 
of an eigenstate 
in terms of properties of a finite set of periodic orbits. 
 
In section 4 we apply the method to the cat map --where the 
semiclassical approximation 
is exact -- and to the baker map where we can test the approximation. 
 
The formulae are here applied as a test to simple maps, but they can be 
easily adapted 
to more realistic Hamiltonian systems by selecting an appropriate 
Poincar\'e section 
\cite{sara-simo}.

\section {Projector on single eigenstates} 
 
For a map, the density of Floquet states can be expressed in the usual 
way in terms of 
traces by the formula 
 
\begin{equation} 
d(\epsilon) = \frac{2\pi}{N}+\sum_{n=1}^{\infty} [tr \hat U^n  e^{- i 
\epsilon n} ] + 
C.C. 
\end{equation} 
$1/N$ plays the role of Plank's constant $h$, $n$ is the discrete time, 
and $\hat U$ is 
the unitary quantization of the map. The density of states results in a 
$2\pi$ periodic 
function with unit $\delta$-spikes at the eigenangles $\epsilon_k$, 
defined by  
 
\begin{equation} 
\hat U | \psi_k \rangle = e^{i \epsilon_k} | \psi_k \rangle \ . 
\end{equation} 
 
Semiclassical approximations using this formula require the calculation 
of traces of 
powers of the map for large values of $n$ which are in turn related to 
classical orbits 
of the map by the Gutzwiller-Tabor \cite{gutzw} trace formula 
 
\begin{equation} 
b_n = tr \hat U^n \approx \sum_{\gamma} A_{\gamma} \exp \left[ 
i(S_{\gamma}/\hbar - 
\mu_{\gamma} \pi /2) \right] \label{gutz} \ , 
\end{equation} 
where $\gamma$ label the $n$-periodic orbits of the map, $S_{\gamma}$ is 
the action, 
$\mu_{\gamma}$ is the Maslov index, and $A_{\gamma}$ is a coefficient 
related to the 
monodromy matrix. The approximation is valid for fixed $n$ in the limit 
as $N 
\rightarrow \infty$ (but not vice-versa). 
 
A more efficient scheme starts from the spectral determinant 
 
\begin{equation} 
P(s) = \det (\hat I - s \hat U) 
\end{equation} 
which vanishes at $s = s_k=e^{ - i \epsilon_k}$ and which can be 
expanded as 
 
\begin{equation} 
P(s) = \sum_{n=0}^N \beta_n s^n \ . 
\end{equation} 
The coefficients $\beta_n$ can be computed recursively from the traces 
as 
 
\begin{equation} 
\beta_n = -\frac{1}{n} \sum_{j=1}^n \beta_{n-j} b_j  \label{betarec} 
\end{equation} 
and therefore when computed semiclassically as in (3) they are expressed in 
terms of linear 
combinations of periodic orbits (pseudo orbits \cite{pseudo}) of period
up to $n$. 
 
The advantage over the calculation in term of traces is that only $N$ 
coefficients need 
to be calculated. Moreover as a consequence of unitarity the symmetry 
 
\begin{equation} 
\beta_{N-j} = (-1)^N \det \hat U ~ \bar{\beta}_j \label{rever} 
\end{equation} 
can be imposed, so that only $N/2$ coefficients are needed and therefore 
only periodic 
trajectories for times up to $N/2$ (i.e. half the Heisenberg time) are 
involved for a 
semiclassical calculation. This is now an optimal encoding of the 
eigenvalues in terms 
of traces, with the $N$ real phases $\epsilon_k$ expressed exactly in 
terms of the $N/2$ 
complex coefficients $\beta_n$ 
 
It is convenient to define another function \cite{bogo2}, proportional 
to the spectral 
determinant, which has the same roots. Utilizing the symmetry 
(\ref{rever}), for an 
even $N$, this function can be written 
 
\begin{equation} 
Z(s) = \frac{P(s)}{\beta_{N/2} ~ s^{N/2}} = 1 + \frac {1}{\beta_{N/2}} 
\sum_{n=0}^{N/2-1} \beta_n ~ s^{n-N/2} + \frac{1}{\bar{\beta}_{N/2}} 
\sum_{n=0}^{N/2-1} 
\bar{\beta}_n ~ s^{N/2-n} \ . 
\end{equation} 
 
When $s$ is on the unit circle $1/s=\bar s$, and then the second sum is 
the complex 
conjugate of the first. Then the function $Z(s)$ becomes real on 
$|s|=1$.

Unitarity also implies that the spectral determinant has all its zeros 
on the unit 
circle. However when the coefficients are computed semiclassically this 
is no longer 
guaranteed. If the symmetry (\ref{rever}) is imposed the resulting 
spectral determinant 
becomes self-inversive, which is a necessary -but by no means 
sufficient- condition for 
unitarity \cite{bbl}. Self-inversive polynomials have the following properties 
\cite{marsden}: 
 
1) All zeros are either on the unit circle or symmetric with respect to 
it. Thus as a 
function of a continuous parameter the zeros of a self-inversive 
polynomial can only 
leave the unit circle in pairs by degenerating on it. 
 
2) If $P(s)$ has only single roots then $dP/ds$ has no zeros on the unit 
circle.

\subsection {Green Operator}

Very similar techniques can be implemented for the calculation of matrix 
elements and 
therefore of eigenfunctions. To obtain the eigenfunctions of the unitary 
matrix $\hat 
U$, we define a Green operator  
 
\begin{equation} 
\hat G (s) = \frac {f(\hat U)}{\hat I - s \hat U} \ , 
\end{equation} 
which, as an analytic function of $s$, has poles at $s = s_k = e^{ - i 
\epsilon_k}$, and 
the residues are proportional to the projectors $\hat P_k = |\psi_k 
\rangle \langle 
\psi_k |$. To obtain normalized projectors, we instead define a {\it 
normalized} Green 
operator 
 
\begin{equation} 
\hat g (s) = \frac {\hat G (s)}{tr[\hat G (s)]} \ . 
\end{equation} 
 
At the poles, the singularities cancel and we obtain $\hat g (s_k) = 
\hat P_k$. $f(\hat 
U)$ is left arbitrary for the moment, and specific choices -made below- 
will be used to 
obtain formulae with different properties. 
 
We can expand the preceding expression utilizing the transpose cofactor 
matrix to write 
the Green operator 
 
\begin{equation} 
\hat g (s) = \frac {f(\hat U) ~ C^t (\hat I - s \hat U)}{ tr[f(\hat U) ~ 
C^t (\hat I - s 
\hat U)]} \label{gs} \ . 
\end{equation} 
The transpose cofactor matrix can be written as a finite power series in 
$s$ 
 
\begin{eqnarray}  
C^t (\hat I - s \hat U) &=& \sum_{n=0}^{N-1} s^n \hat X _n \ , \\ 
\hat X _n &=& \sum_{i=0}^n \beta_i ~ \hat U ^{n-i} \label{xn} \ . 
\end{eqnarray} 
 
By the unitarity of $\hat U$, the $\hat X _n$ matrices satisfy the 
following symmetry 
 
\begin{equation} 
\hat X _{N-j} = - \hat U^{\dagger} (-1)^N \det \hat U ~ \hat X 
_{j-1}^{\dagger} \ . 
\label{reverx}
\end{equation}

Utilizing this symmetry two choices of $f(\hat U)$ lead to expansions 
with useful 
properties: 
 
\noindent a) For $f(\hat U)= \hat U$, for even $N$, the corresponding 
Green function is 
 
\begin{equation} 
\hat g_a (s) = \frac{ \sum_{n=0}^{N/2-1} s^n ~ \hat U \hat X_n - \det 
\hat U ~ s^{N-1} 
\sum_{n=0}^{N/2-1} s^{-n} ~ \hat X^{\dagger}_n } { \sum_{n=0}^{N/2-1} 
s^n ~ tr 
\left(\hat U \hat X_n \right) - \det \hat U ~ s^{N-1} \sum_{n=0}^{N/2-1} 
s^{-n} ~ tr 
\left(\hat X^{\dagger}_n \right)}  \ . 
\end{equation} 
 
Replacing the expansion of the $\hat X_n$ operators in powers of $\hat 
U$, we obtain 
 
\begin{equation} 
\hat g_a(s) = \frac{\sum_{i=0}^{N/2-1} c_i (s) ~ \hat U ^{i+1} - \det 
\hat U ~ s^{N-1} 
\sum_{i=0}^{N/2-1} \bar{c_i} (1/\bar s) ~ \hat U^{\dagger ~ i} }  
{\sum_{i=0}^{N/2-1} 
c_i (s) ~ tr \left(\hat U ^{i+1} \right) - \det \hat U ~ s^{N-1} 
\sum_{i=0}^{N/2-1} 
\bar{c_i} (1/\bar s) ~ tr \left(\hat U^{\dagger ~ i} \right) }  
\label{greenfinal} \ , 
\end{equation} 
in which the projector is expressed as a finite combination of the first 
$N/2$ powers of 
the propagator. 
 
The coefficients $c_i$ are obtained from the $\beta_n$ 
 
\begin{equation} 
c_i (s) = \sum_{n=i}^{N/2-1} \beta_{n-i} ~ s^n \label{cis} \ . 
\end{equation} 
 
This normalized Green operator has no singularities as $s$ goes around 
on the unit 
circle. This can be demonstrated because the denominator is equal to  
$-dP/ds$ (see 
Appendix A). Then by the second property of the self-inversive 
polynomials it is 
guaranteed that this denominator never takes the zero value for $s$ over 
the unit 
circle. This property, guaranteed by unitarity, will remain true --even 
when the traces 
and propagator are calculated semiclassically-- as long as self 
inversiveness is 
maintained.

\noindent b) For $f(\hat U)= \hat I + s \hat U$ , $\hat G(s)$ in (9) is 
the   
Cayley transform of $\hat U$ and is therefore hermitian on the unit 
circle. In 
this case 
\begin{equation} 
\hat g_b (s) = \frac{\sum_{n=0}^N s^n ~ \hat Y_n}{\sum_{n=0}^N s^n ~ tr 
\left(\hat Y_n 
\right) } 
\end{equation} 
with 
\begin{equation} 
\hat Y_n = \hat X_n + \hat U \hat X_{n-1} = 2 \hat X_n - \beta_n \hat I 
\label{yn} \ . 
\end{equation} 
These new operators satisfy a simpler symmetry relation than the 
preceding $\hat X_n$, 
analogous to the symmetry of the $\beta_n$ coefficients 
 
\begin{equation} 
\hat Y_{N-j} = - (-1)^N \det \hat U ~ \hat Y^{\dagger}_j \ . 
\label{revery}
\end{equation} 
In this case, for even $N$, the normalized Green operator can be 
written as 
 
\begin{equation} 
\hat g_b(s) = \frac{\hat{\Sigma} (s) - \hat{\Sigma}^{\dagger} (1/\bar 
s)} {tr\left[ 
\hat{\Sigma} (s) \right] - tr \left[ \hat{\Sigma}^{\dagger} (1/\bar s) 
\right]} 
\label{gb} \ , 
\end{equation} 
where 
 
\begin{equation} 
\hat{\Sigma} (s) = \frac{1}{\beta_{N/2}} \left[ \frac{1}{2} \hat Y_{N/2} 
+ 
\sum_{n=0}^{N/2-1} s^{n-N/2} ~ \hat Y_n  \right] \ . 
\end{equation} 
 
In this expression, more symmetrical than (\ref{greenfinal}), the 
numerator is 
anti-hermitian on the unit circle, and the denominator is purely 
imaginary. The basic 
operator $\hat{\Sigma} (s)$ is a Fourier transform of the $\hat Y_n$ 
operators and their 
traces. When utilizing this finite representation, they then become the 
basic objects of 
study in the time domain, in place of the propagator and its traces. 
 
When $\hat U$ is a finite matrix, $\hat g_a(s)$ and $\hat g_b(s)$ are 
both rational 
analytic functions of $s$. They are identical at the eigenvalues $s=s_k$ 
(on the unit 
circle). However their zero and singularity structure is quite 
different. 
 
In particular $\hat g_a(s)$, as noted, has a denominator which does not 
vanish on the 
unit circle, even when the coefficients are computed approximately, as 
long as 
self-inversiveness is imposed on them. The operator $\hat g_a(s)$ then, 
has no 
singularities in a small strip enclosing the unit circle and can be 
followed continuously 
from eigenvalue to eigenvalue. A disadvantage is that $\hat g_a(s)$ is 
not hermitian 
(except at the values $s=s_k$). On the contrary $\hat g_b(s)$ is 
hermitian on the unit 
circle, but its singularities (the zeroes of the denominator) are 
precisely on the unit 
circle, in between two successive eigenvalues. 
 
It should be noticed that, although for simplicity all the above formulae
are written for even $N$, they are easily adapted to the odd case, with
minimal changes.  

 The preceding operator expressions for $\hat g_a(s)$ and $\hat g_b(s)$ 
provide a common structure that underlies all further semiclassical
approximations. Depending on the representation used to evaluate them, 
they can yield results for transition probabilities, phase space 
distributions, response functions or wave function correlations. This is 
achieved by linking the spectral properties of the map to a {\it finite 
number } of traces and  matrix elements of the propagator. To apply them
one needs the semiclassical evaluation of the propagator traces as in
(\ref{gutz})- which are representation independent- and also of the 
semiclassical propagator in the chosen representation.
For example if they are computed in the coordinate representation they 
provide  expressions for $|\langle x| \psi_k \rangle |^2$. If the propagator 
is computed in the Weyl representation \cite{ozorio} they yield the Wigner 
distribution of eigenfunctions \cite{fishman}. 
In all cases the unitarity of the propagator - which is generally only 
approximate in semiclassical treatments - is partially restored by the 
imposition of the symmetries (\ref{rever},\ref{reverx},\ref{revery}).
In the next section the above formalism is developed in the coherent state 
representation to obtain expressions for semiclassical Husimi distributions.

\section{Semiclassical approximation in the coherent state representation} 
 
The formulae derived in the previous section are exact and for maps can, 
at first sight, 
seem just a complicated way of looking at a simple algebraic eigenvalue 
problem. 
However, they are prepared in such a way as to make the semiclassical 
transition as 
clear as possible and to unveil the necessary approximations involved in 
such a 
transition,  
 
The coherent state mean value of $\hat g(s)$ is 
 
\begin{equation} 
{\cal H} (s, z, \bar z) \equiv \frac{\langle z | \hat g(s) | z 
\rangle}{\langle z | z \rangle} \ . 
\end{equation} 
This is a phase space distribution, analytic in $s$, which becomes the 
Husimi 
distribution of the eigenstates $|\psi_k \rangle$ at the values $s_k = 
e^{- i 
\epsilon_k}$.  
 
Its properties, reflecting the zeroes and singularities of $\hat g(s)$ 
can be followed 
continuously in the $s$ complex plane. Its calculation involves the $N/2$ 
coherent state return amplitudes 
 
\begin{equation} 
C_n (z, \bar z) \equiv \frac{\langle z | \hat U^n | z \rangle}{\langle z 
| z \rangle} 
\end{equation} 
in addition of the $N/2$ traces 
 
\begin{equation} 
b_n=tr \hat U^n = \int dz ~ d\bar z ~ C_n (z,\bar z) \ . 
\end{equation}  

The first admits immediately a well known semiclassical approximation in 
the time 
domain, namely the Gutzwiller-Tabor trace formula (\ref{gutz}). The 
stationary phase 
approximations involved in the derivation are of course valid for fixed 
time $n$ and for 
$\hbar \rightarrow 0$, or equivalently, $N \rightarrow \infty$, and 
their use in the 
regime $n \approx N/2$ needed for the exact calculation is questionable. 
 
The semiclassical calculation of the return amplitudes $C_n (z, \bar z)$ can 
be done by stationary phase calculation of the integral 
 
\begin{equation} 
\langle z_f| \hat U^n |z_i \rangle = \int \int \langle z_f | q_f \rangle 
\langle q_f | 
\hat U^n | q_i \rangle \langle q_i | z_i \rangle ~ dq_i ~ dq_f 
\label{intzq} \ , 
\end{equation} 
where 
\begin{equation} 
\langle z| q \rangle = \frac{\exp \left\{ -\frac{1}{2 \hbar} \left[ z^2 
+ q^2 - 2 \sqrt 
2 z q \right] \right\} }{\sqrt [4]{\pi \hbar}} 
\end{equation} 
is the coherent state in coordinate representation, and the 
propagator has the usual Van Vleck form 
\begin{equation} 
\langle q_f| \hat U^n |q_i \rangle = \frac{1}{\sqrt{2\pi i \hbar}} 
\sum_{\gamma} 
\sqrt{\frac{\partial ^2 S^{(n)}_{\gamma} (q_i,q_f) }{\partial q_i 
\partial q_f}} ~ \exp \left[ 
~ \frac{i}{\hbar} S^{(n)}_{\gamma} (q_i,q_f) \right] \label{vleck} \ . 
\end{equation} 
The sum must be made over all the paths in the phase space of the 
corresponding 
classical map that connect $q_i$ with $q_f$ in $n$ steps. Additional phases 
connecting the different branches, the Maslov indices, must be added to 
$S^{(n)}_{\gamma}$. 
 
The calculation of the integral (\ref{intzq}) with the Van Vleck 
approximation 
(\ref{vleck}) can be done as usual by stationary phase and the result 
has the structure 
of another Van Vleck formula with complex coordinates for a generating 
function. 
 
\begin{equation} 
\langle z_f | \hat U^n |z_i \rangle = \sum  \sqrt{ -i \frac{\partial ^2 
G_3}{\partial 
\bar z_i \partial z_f} } ~ \exp \left[ -\frac{i}{\hbar} G_3(\bar 
z_i,z_f) \right] 
\label{zuzsem} \ . 
\end{equation} 
 
This new classical generating function is a complex Legendre transform 
of the action $S$ 
 
\begin{equation} 
G_3 = -S + \frac{1}{2} (p_f q_f - p_i q_i) + \frac{i}{4} (q_f^2 + p_f^2 
+ q_i^2 + p_i^2) 
\label{gzz} \ , 
\end{equation} 
or equivalently, in the complex coordinates 
 
\begin{eqnarray} 
z &=& \frac{1}{\sqrt{2}} \left(q - i p \right)  \\ 
\bar z &=& \frac{1}{\sqrt{2}} \left(q + i p \right) \ ,   
\end{eqnarray} 
it can be written as 
 
\begin{eqnarray} 
G_3 =  -S + \frac{i}{4} \left(z_f^2 + 2 z_f \bar z_f - \bar z_f^2 
\right) - \frac{i}{4} 
\left( z_i^2 - 2 z_i \bar z_i - \bar z_i^2 \right) \label{g1z} \ . 
\end{eqnarray} 
 
This function depends explicitly on $\bar z_i,z_f$ variables, and yields 
the equations 
of motion 
 
\begin{equation} 
z_i = -i \frac{\partial G_3}{\partial \bar z_i} \ , ~~~~~~~~~  \bar z_f = -i 
\frac{\partial 
G_3}{\partial z_f} \label{motion} \ . 
\end{equation} 
 
These are algebraic equations that define the map implicitly provided 
$\frac{\partial^2 
G_3}{\partial \bar z_i \partial z_f} \neq 0$ where the sum must be made 
over complex 
paths that connect the point $\bar z_i$ to $z_f$. These formulae show 
that the return amplitude $C_n$ has as exponent the function 
$G_3(\bar z,z) -i \bar z z$ which using 
(\ref{motion}) can be shown to be stationary at the periodic points. 
Quadratic approximation of the generating function at these points yields 
the coherent state return amplitudes as a sum of Gaussian peaks centered 
on them 
 
\begin{equation} 
C_n(z,\bar z) = \sum_{\gamma} \frac{1}{\sqrt{u_{\gamma}}} \exp \left[ - 
\frac{1}{\hbar} {\cal G}^{\gamma} (z,\bar z) + \frac{i}{\hbar} S_{\gamma} 
\right] \label{cn} \ . 
\end{equation} 
 
\begin{equation} 
{\cal G}^{\gamma} (z,\bar z) = \frac{1}{u_{\gamma}} \left[ \frac{\bar 
v_{\gamma}}{2} 
(\bar z - \bar z_{\gamma})  ^2 + (u_{\gamma}-1) (\bar z - \bar 
z_{\gamma}) (z - 
z_{\gamma}) - \frac{v_{\gamma}}{2} (z - z_{\gamma})^2 \right] 
\label{gexp} \ ,  
\end{equation} 
where $z_{\gamma},\bar z_{\gamma}$ are the position of the periodic 
points, $S_{\gamma}$ 
are the corresponding actions, and $u_{\gamma},v_{\gamma}$ and $\bar 
v_{\gamma}$ are the 
elements of the complex monodromy matrix defined by 
 
\begin{equation} 
\left(  
\begin{array}{cc} 
\partial z_f / \partial z_i & \partial z _f / \partial \bar z_i  \\ 
\partial \bar z_f / \partial z_i & \partial \bar z_f / \partial \bar z_i  
\end{array} 
\right) = \left(  
\begin{array}{cc} 
u & \bar v   \\ 
v & \bar u  
\end{array} 
\right) 
\end{equation} 
calculated on the corresponding periodic point. 
 
Using (\ref{cn}) for the return amplitudes and (\ref{gutz}) for the 
traces, the 
Husimi distributions of eigenstates are explicitly written in terms of 
properties of 
periodic points of all periods up to half the Heisenberg time $N/2$. 
 
Although the semiclassical approximations are available in the time 
domain for the 
traces $b_n$ and the return amplitudes $C_n(z,\bar z)$ it is clear from 
our formulae 
that the essential ingredients for the calculations are the coefficients 
$\beta_n$ and 
the operators $\hat X_n$ (or $\hat Y_n$). These are related by 
(\ref{betarec}) and 
(\ref{xn}) to $b_n$ and $C_n$ and therefore can be recursively computed 
from them. This 
procedure yields the $\beta_n$ as linear combinations of {\it products} 
--the so called 
pseudo-orbits-- of terms involving many periodic orbits of different 
periods and with 
signs implying many possible cancellations between long and short 
orbits. The same 
situation prevails for the $\hat X_n$ (or $\hat Y_n$). The expressions 
yield linear 
combinations of return amplitudes at many different times. In fact $\hat 
X_n(z,\bar z)$ 
give the phase space distribution of pseudo-orbits and it is easy to 
show that 
$\beta_n=(N-n) ~ tr \hat X_n$. In view of the cancellations involved it 
would be very 
desirable to have a direct semiclassical derivation --both for $\beta_n$ 
and $\hat 
X_n$-- that would not go through the indirect procedure of approximating 
$b_n$ and $\hat 
U_n$, and then recursively computing $\beta_n$ and $\hat X_n$. However 
to our knowledge 
such a derivation is not available.

\section{Numerical calculations}

We will test the methods proposed on two simple well known maps: the cat 
map, and the 
baker map. 
 
For the cat map 
 
\begin{eqnarray} 
\left(  
\begin{array}{c} 
q_f  \\ 
p_f  
\end{array} 
\right) = \left(  
\begin{array}{cc} 
a & b   \\ 
c & d 
\end{array} 
\right)  \left(  
\begin{array}{c} 
q_i  \\ 
p_i 
\end{array} 
\right)  (mod ~ 1) 
\end{eqnarray} 
with $a,b,c$ and $d$ integer numbers, and $ad-bc=1$, the Van Vleck 
formula in the 
$q_i,q_f$ representation is exact \cite{hannay} and the corresponding 
one in the 
coherent state representation can be computed explicitly. 
 
The transformation to complex coordinates leads to the map 
 
\begin{eqnarray} 
\left(  
\begin{array}{c} 
z_f  \\ 
\bar z_f  
\end{array} 
\right) = \left(  
\begin{array}{cc} 
u & \bar v   \\ 
v & \bar u 
\end{array} 
\right)  \left(  
\begin{array}{c} 
z_i  \\ 
\bar z_i 
\end{array} 
\right)  
\end{eqnarray} 
with 
 
\begin{eqnarray} 
u &=& \frac{1}{2} \left [ (a+d) + i(b-c) \right ] \ , \\ 
v &=& \frac{1}{2} \left [ (a-d) + i(b+c) \right ] \ . 
\end{eqnarray} 
and its quantization with (\ref{vleck}) is obtained exactly by Gaussian 
integration. The coherent state return amplitudes are computed with the 
formulae (\ref{cn}) and (\ref{gexp}), with the 
advantage that the coefficients of the monodromy matrix are the same for 
all the 
periodic points of a given period. Since the cat map is defined on the 
torus we use 
periodized coherent states \cite{voros}. With this result we can test 
our formulae 
keeping in mind that for this special case the semiclassical 
approximation is exact. 
 
In Fig.1 we first look at quantities in the time domain, i.e. as a 
function of $n$. The 
correlations $|C_n(z, \bar z)|^2$ display for short times Gaussian peaks 
at the 
n-periodic points. 
 
However we have seen that the irreducible information about eigenstates 
is carried by 
more complicated operators such as the $\hat X_n$ in (\ref{xn}) or the 
$\hat Y_n$ in 
(\ref{yn}). These operators carry the phase space information that 
corresponds to the 
pseudo-orbits in the resummation of trace formula. In fact many 
different powers of 
$\hat U$ contribute to them and give rise to complex interference 
structures in phase 
space. The pseudo-orbit contributions --contained in the coefficient 
$\beta_n$-- are 
obtained from the fact that $\beta_n = (N-n) ~ tr \hat X_n$. 
 
In Fig.2 we have displayed the first few operators $\hat U \hat X_n$ 
appearing in the 
expansion of $g_a(s)$. For very short times we notice the simple 
superposition of the 
patterns of $C_n(z,\bar z)$. In particular $\hat U \hat X_2$ shows peaks 
at the period 
1, period 2 and period 3 periodic orbits. For longer times ($\hat U \hat 
X_3$) the 
interferences between orbits of different periods are very strong and 
one can barely 
notice the period 4 orbits, while the dominant structure is of period 1. 
This is in 
striking contrast with the regular appearance of periodic points in the 
return amplitudes. Notice also the totally unexpected --and for us 
unexplained-- appearance of a strong  
periodicity (of period three) in the phase space patterns, which is 
totally invisible in the $C_n$ 
coefficients. We cannot give an explanation to this behavior except 
noting that a direct 
semiclassical calculation of these coefficients would be very desirable 
and might be 
more effective than one which goes through the calculation with periodic 
points. 
 
The smooth dependence of the operator $\hat g(s)$ on the quasi energy 
variable allow us 
to consider the exact response function as a distribution in phase space 
depending 
continuously on $s$, thus allowing an {\it exact} interpolation for all 
values of $s$. 
The Husimi eigen-distributions are particular values of this function 
which 
should be positive and display a pattern of $N$ zero values, in 
accordance 
with the general properties of Husimi distributions of pure states on 
the 
torus. \cite{voros} 
 
We display in Fig.3 the real part of ${\cal H}_a (e^{-i \epsilon}, z, 
\bar z)$ for 
values of $\epsilon$ near an eigenvalue. The distribution shows a very 
stable pattern of 
minima that gradually develops into $N=14$ zeros at 
$\nu=\epsilon/2\pi=0.4167$. At this 
point $Z(e^{-i \epsilon})=0$, the distribution is positive and we 
verify that the 
imaginary part vanishes for every phase space point. This is the signal 
\cite{voros} 
that a true projector has been reached and that the Husimi distribution 
is the modulus 
of an analytic function with $N$ zeros. Past the eigenvalue the 
distribution can become
negative but the pattern of minima remains very stable. At the values 
where $Z(e^{-i 
\epsilon})$ has an extremum, in between zeros, we show in the appendix 
that the 
distribution becomes flat with value $1/N$. At this point the pattern of 
minima changes 
and picks up the structure of zeros of the next eigenfunction. 
 
For cat maps the semiclassical formulae are exact and therefore the above 
results merely confirm that the whole scheme is consistent and accurate but
they do not really test the semiclassical approximation. 
Towards this purpose we consider another simple linear map --the baker's-- 
where the semiclassics is not exact. The classical $T$-iteration of the 
baker map from the $(q_0, p_0)$ to $(q_T, p_T)$ real coordinates 
\cite{baker}, takes the complex form
 
\begin{eqnarray} 
z_T &=& u_T ~ z_0 + v_T ~ \bar {z_0} - w_T \\ 
\bar z_T &=& v_T ~ z_0 + u_T ~ \bar {z_0} - \bar w_T 
\end{eqnarray} 
 
\begin{eqnarray} 
u_T &=& \frac{2^T + 2^{-T}}{2} = \cosh (T \ln 2) \\ 
v_T &=& \frac{2^T - 2^{-T}}{2} = \sinh (T \ln 2) \\ 
w_T &=& \frac{1}{\sqrt{2}} \left( \gamma + i ~ 2^{-T} \gamma ^{\dagger} 
\right) 
\end{eqnarray} 
$\gamma$ is the integer part of $2^T q_0$ and $\gamma^{\dagger}$ is the number 
obtained after 
inversion of their binary digits. Each possible $\gamma$ defines a periodic 
point $z_{\gamma}=(\gamma-i\gamma^{\dagger})/\sqrt{2}(2^T-1)$. The return 
amplitudes $C_n (z, \bar z)$ are expressed as a sum over them 
 
\begin{equation} 
C_n(z,\bar z) = \frac{1}{\sqrt{u_T}}  \sum_{\gamma=1}^{2^T-1} \exp \left[ - 
\frac{1}{\hbar} {\cal 
G}^{\gamma} (z,\bar z) +  \frac{i}{\hbar} S_{\gamma} \right] \ , 
\end{equation} 
where the complex generating function takes the standard form 
\begin{equation} 
{\cal G}^{\gamma} (z,\bar z) = \frac{1}{u_T} \left[ \frac{v_T}{2} (\bar z - 
\bar z_{\gamma}) 
^2 + (u_T-1) (\bar z - \bar z_{\gamma}) (z - z_{\gamma}) - \frac{v_T}{2} (z - 
z_{\gamma})^2 
\right]  
\end{equation} 
and  
\begin{equation} 
S_{\gamma} = \frac{\gamma ~ \gamma ^{\dagger}}{2^T-1} 
\end{equation} 
is the action of the corresponding periodic orbit. 
 
In Fig.4 we show the exact passage of the distribution through an 
eigenvalue ($N=16$) where - as expected - the same general pattern as the 
previous example  can be observed. 

In Figs.5 and 6 we test the semiclassical
approximation by comparing both the spectrum and the eigendistributions with
the exact results. Fig.5 shows the $Z$ function for a small stretch of the 
unit circle comprising three zeroes ($ N=20 $) and showing that the 
semiclassical zeroes are approximated well within a fraction of the mean 
eigenvalue separation. 
 In Fig.6 we show the three eigendistributions: the top three are the
 exact ones obtained at the exact zeroes, while the bottom three are
 the corresponding semiclassical ones (computed at the semiclassical zeroes).
 Considering that the distributions are obtained by combining the classsical
 information of $ 2^{N/2}\approx 1000 $ periodic points, it is remarkable 
how well the approximation captures the essential structure
 of the eigenfunctions. One important point should be noticed: in the 
semiclassical case there is no guarantee that the distribution calculated at
 a zero of $Z$ will correspond exactly to a one dimensional projector 
(i.e. to a pure state) and therefore its Husimi distribution may not have 
exactly $N$  zeroes and may be even negative. This is clearly the case in 
the first semiclassical state at $\nu=0.4734$. The large white area has 
negative values - and is not contoured - but corresponds very well with 
the central area where the zeroes of the exact distribution are located. 
The positive top of the distribution is roughly well located but it looks 
like the whole distribution
 had been shifted towards negative values, while mantaining a very accurate
 correspondence of the maxima and minima. The second state at $\nu=0.5445$
 reproduces very well the top part of the distribution, while now the minima
 have become positive, but again maintaining a very accurate correspondence
 with the exact pattern of zeroes.  The third state at $\nu=0.6088$ 
reproduces in very fine detail the patterns of the exact state. Overall 
we have observed that this state of affairs is quite typical: the position 
of the maxima and minima of the semiclassical states are very well 
reproduced but the relative heights of peaks and valleys are not always 
accurate.

\section{Conclusions} 
We have presented a method that can, in principle, provide a reliable 
way to compute 
single Husimi eigendistributions and which has the same advantages --and 
drawbacks-- as 
the resummation methods based on the spectral determinant for the 
calculation of single 
eigenvalues. The final results (\ref{greenfinal}) and (\ref{gb}) for the 
Green operators 
when approximated semiclassically give approximations to the projectors 
on single eigenstates as sums of 
Gaussian centered on periodic points with periods up to half the 
Heisenberg time. Similar 
expressions, with contributions not so well localized, can be obtained 
for Wigner 
functions or probability distributions. The usual exponential 
proliferation of periodic 
orbits in a chaotic system is the main obstacle to applications in the 
deep semiclassical 
regime. These are however not limited to simple maps as the method has 
been applied also 
to the Bunimovich stadium in \cite{sara-simo}. 
 
We acknowledge many important discussions with F. Simonotti and P. Leboeuf, 
and partial support from ANPCyT program PICT97-01015 and ECOS-Secyt A98E03.

\appendix 
\section{Normalized Green operator at $Z$-function extremes}

To calculate the denominator of $\hat g_a$ we use a well known property 
of the 
parametrized matrices  
 
\begin{equation} 
\frac{d}{ds} \det \left[ \hat M(s) \right] = \det \left[ \hat M(s) 
\right] tr \left[ 
\frac{d}{ds} \left[ \hat M(s) \right] \hat M^{-1} (s) \right] \ . 
\end{equation} 
 
If we use this property for $\hat M(s)=\hat I - s \hat U$, then its 
determinant is the 
characteristic polynomial, and we have 
 
\begin{equation} 
\frac{d}{ds} P(s) = P(s) ~ tr \left[ - \hat U (\hat I - s \hat U)^{-1} 
\right] \ . 
\end{equation} 
Then if $P(s)$ is introduced into the trace, we obtain the transpose 
cofactor matrix, 
and we demonstrate 
 
\begin{equation} 
tr \left[ \hat U ~ C^t \left( \hat I - s \hat U \right) \right] = - 
\frac{d}{ds} P(s) \ , 
\end{equation} 
that it is exactly the denominator of $\hat g_a$. Recalling the definition 
of $Z(s)$ function 
 
\begin{equation} 
Z(s) = \frac{P(s)}{\beta_{N/2} s^{N/2}} \ ,  
\end{equation} 
we calculate its derivative respect to the real parameter $\epsilon$ 
that rounds the 
unit circle $(s=e^{-i \epsilon})$ 
 
\begin{equation} 
\frac{dZ}{d \epsilon} =  \frac{dZ}{ds} \frac{ds}{d \epsilon} = 
\frac{i~s^{-N/2}}{ 
\beta_{N/2} } \left[ \frac{N}{2} P(s) - s \frac{d}{ds} P(s)  \right] \ . 
\end{equation} 
Then at $s_{\alpha}=e^{-i \epsilon_{\alpha}}$ where $Z$, as a function 
of $\epsilon$, 
reach a maximum or a minimum, we have the following identity 
 
\begin{equation} 
\frac{dP}{ds} (s_{\alpha}) = \frac{N}{2 s_{\alpha}} P(s_{\alpha}) \ . 
\end{equation} 
 
The denominator of the $\hat g_a$ normalized Green operator is equal to 
the derivative 
of the spectral determinant. Then, in a diagonal basis of $\hat U$, 
$\hat g_a$ 
specialized at a extreme of the $Z$ function is 
 
\begin{eqnarray} 
\hat g_a(s_{\alpha}) &=&  \frac{ \sum_{n=1}^N (1/s_n)  \prod_{k \neq n} 
(1-s_{\alpha} 
/s_k) |\psi_n \rangle \langle \psi_n| } {-\frac{dP}{ds} (s_{\alpha})} 
\nonumber \\ 
&=& - \frac{ \sum_{n=1}^N (1/s_n)  \prod_{k \neq n} (1-s_{\alpha}/s_k) 
|\psi_n \rangle 
\langle \psi_n| } { \frac{N}{2s_{\alpha}} P(s_{\alpha})} \nonumber \\ 
&=& - \frac{2}{N} \frac{s_{\alpha}}{\prod_{k} (1-s_{\alpha} / s_k)} 
\sum_{n=1}^N (1/s_n) 
\prod_{k \neq n} (1-s_{\alpha}/s_k) |\psi_n \rangle \langle \psi_n| \ . 
\end{eqnarray} 
Introducing the spectral determinant of the denominator into the sum we 
obtain 
 
\begin{equation} 
\hat g_a(s_{\alpha}) = - \frac{2}{N} \sum_{n=1}^N \frac{(s_{\alpha} 
/s_n)}{1-(s_{\alpha}/ s_n) } |\psi_n \rangle \langle \psi_n| \ . 
\end{equation} 
 
We can separate the real and the imaginary part of the prefactor, 
knowing that both 
$s_{\alpha}$ and $s_n$ are on the unit circle. 
 
\begin{equation} 
\frac{(s_{\alpha}/s_n)}{1-s_{\alpha} /s_n} = -\frac{1}{2} + \frac{i}{2} 
\frac{ \Im 
(s_{\alpha}/s_n) }{\left[ 1- \Re (s_{\alpha}/s_n) \right]} \ . 
\end{equation} 
 
Contrary to the imaginary part, we see that the real part is independent 
of the index $n$ or $\alpha$. Then the 
normalized Green operator at the extremes of the $Z$ function can be 
written as 
 
\begin{eqnarray} 
\hat g(s_{\alpha}) &=& - \frac{2}{N} \sum_{n=1}^N \left( -\frac{1}{2} + 
\frac{i}{2} 
B_{{\alpha} n} \right)  |\psi_n \rangle \langle \psi_n| \\ 
&=& \frac{1}{N} \sum_{n=1}^N  |\psi_n \rangle \langle \psi_n|  - 
\frac{i}{N} 
\sum_{n=1}^N B_{{\alpha} n} |\psi_n \rangle \langle \psi_n| 
\end{eqnarray} 
and splits into an hermitian and antihermitian part. We have demonstrated 
here that the 
hermitian part is exactly the identity operator over $N$. Then, the real 
part of the 
expectation value of $\hat g(s_{\alpha})$ will be always $1/N$, 
independent of the 
quantum state.

\newpage 
 
\begin{figure} 
\centerline{\psfig{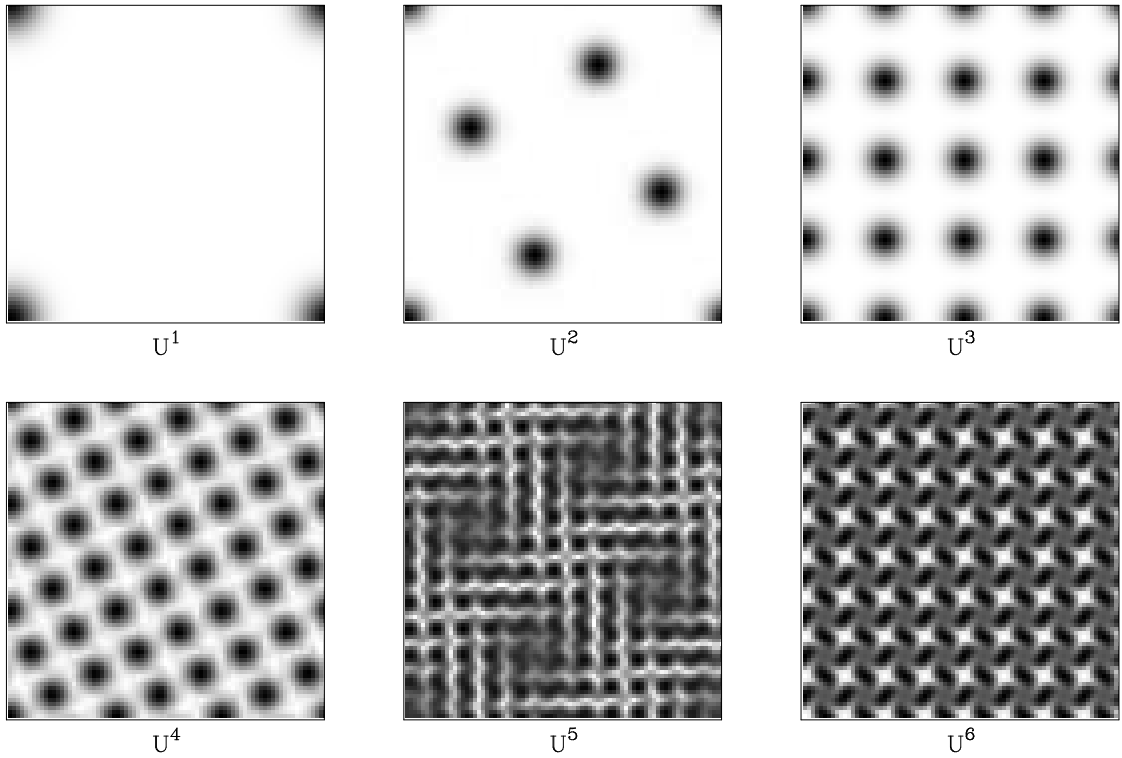}} 
\vspace{0.2cm} 
\caption{\label{ucat} Modulo of the coherent state return amplitude for the 
first six powers of the Arnold cat map propagator. The torus is quantized 
with periodic boundary conditions, and $N=128$. Linear grey scale between 
0 (white), and the maximum (black) of each picture.} 
\end{figure} 
\begin{figure} 
\centerline{\psfig{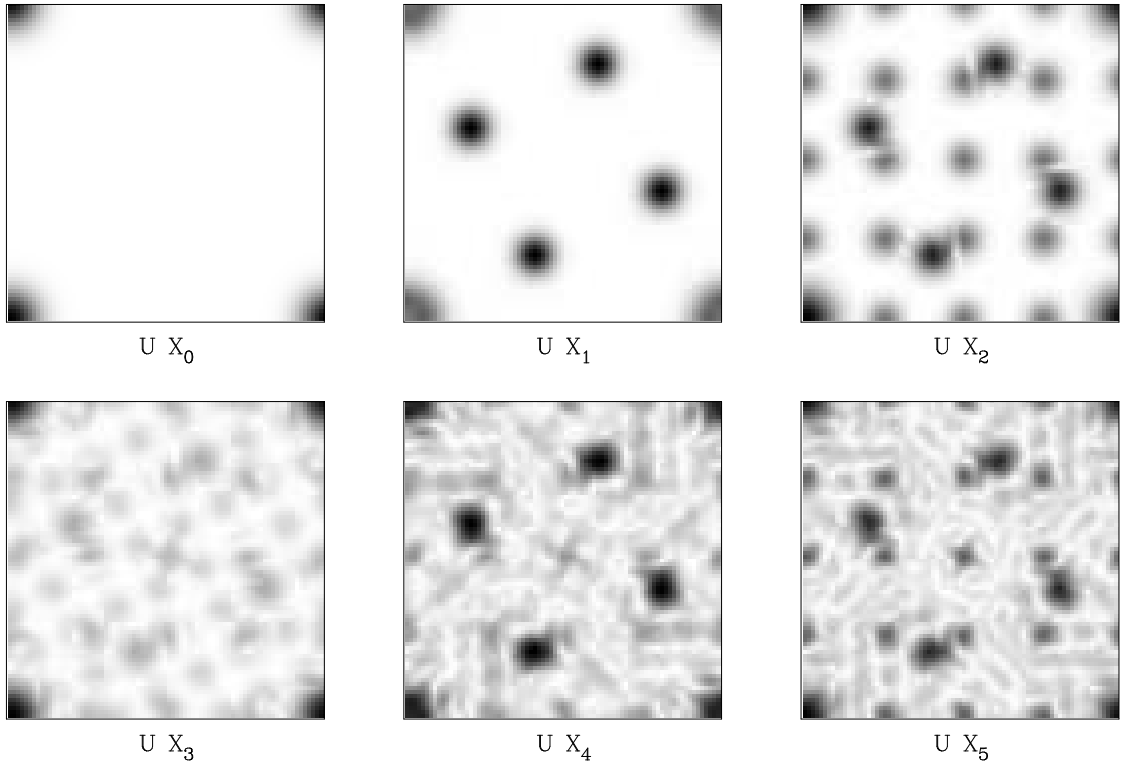}} 
\vspace{0.2cm} 
\caption{\label{uxcat} Modulo of the coherent state mean value for the 
$\hat U \hat X_n$ operators. Same parameters and scales as Fig. 1} 
\end{figure} 
\begin{figure} 
\centerline{\psfig{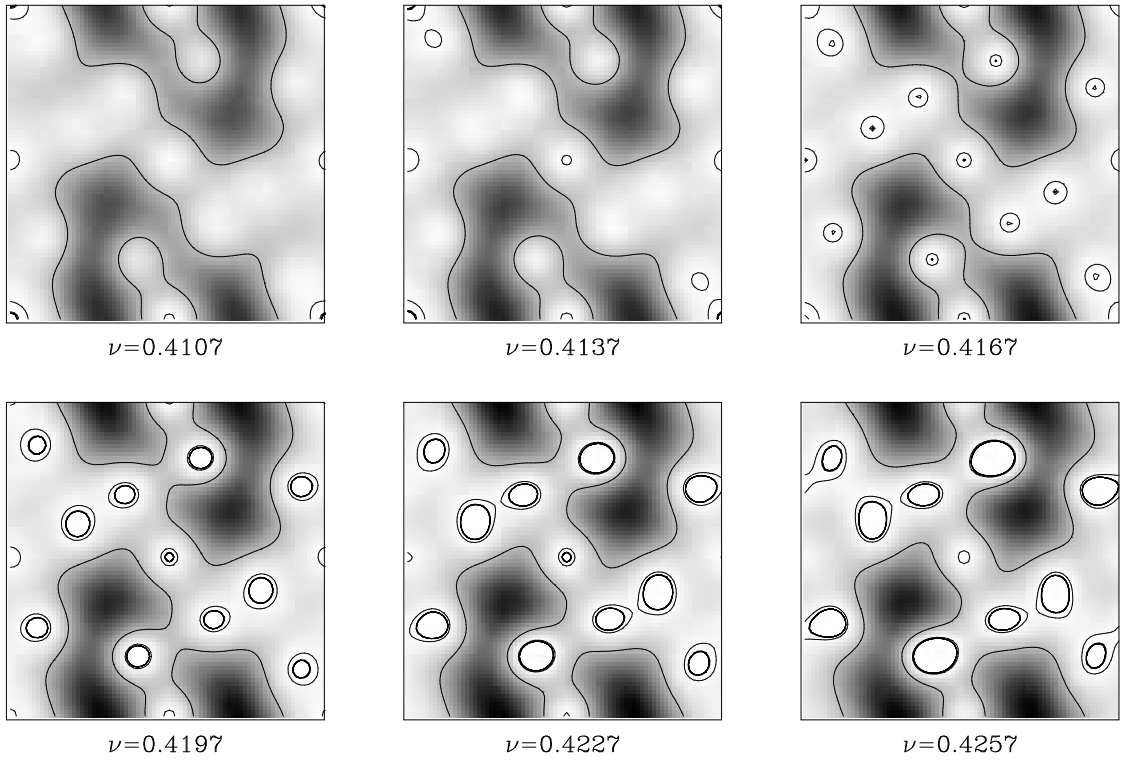}} 
\vspace{0.2cm} 
\caption{\label{gcat} Real part of ${\cal H}_a (s,z,\bar z)$, for $s = \exp 
(-i 2 \pi \nu)$. 
Linear grey scale between 0 (white), and $1/\sqrt{N}$ (black). 
Logarithmic level curves 
at $1/N, 1/N^2,1/N^3,...$. Arnold cat map with periodic boundary 
conditions, and $N=14$. 
There is an eigenvalue of the propagator near $\nu=0.4167$ (3rd. 
picture). The exact and 
semiclassical computation gives the same values.} 
\end{figure} 
\begin{figure} 
\centerline{\psfig{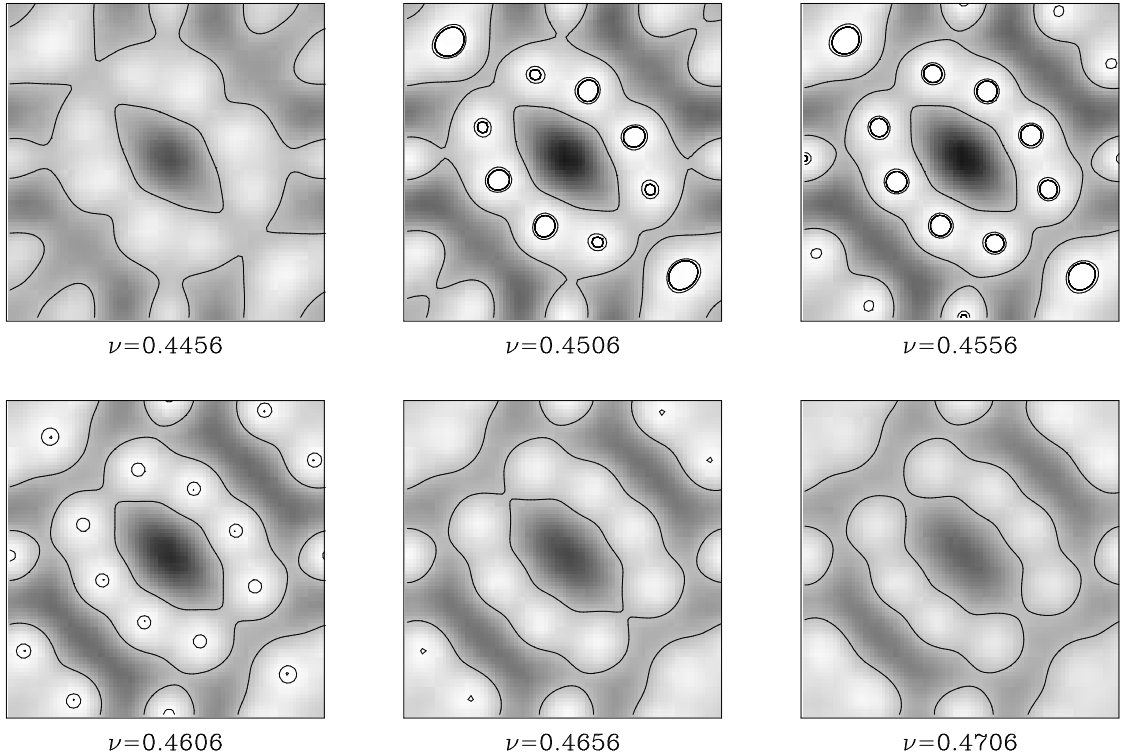}} 
\vspace{0.2cm} 
\caption{\label{gbak} The same as Fig. 3, but for the baker map, with 
anti-periodic boundary 
conditions, and $N=16$. There is an eigenvalue of the propagator near 
$\nu=0.4606$ (4th. 
picture).} 
\end{figure} 
\begin{figure} 
\centerline{\psfig{figure=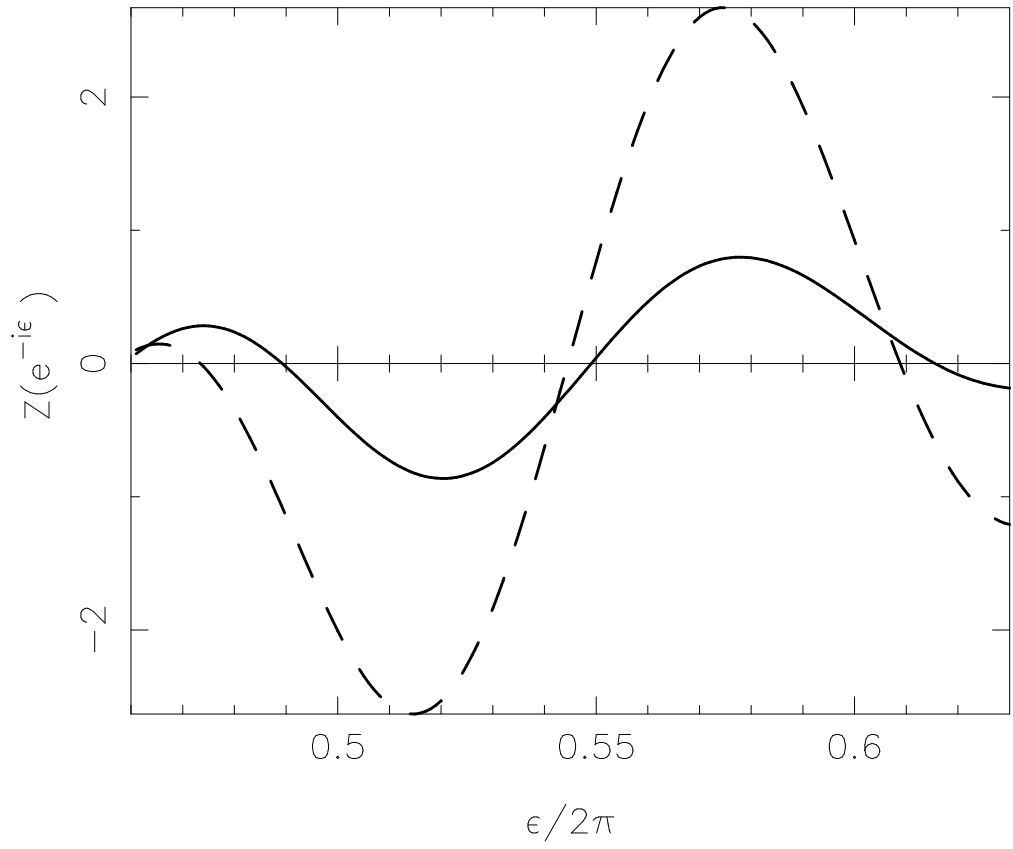}} 
\vspace{0.2cm} 
\caption{\label{zexsem} Exact $Z$ function (bold line), and its semiclassical 
approximation (dashed line), for the baker map, with anti-periodic boundary 
conditions, and $N=20$. The zeroes correspond to the eigenvalues}
\end{figure} 
\begin{figure} 
\centerline{\psfig{figure=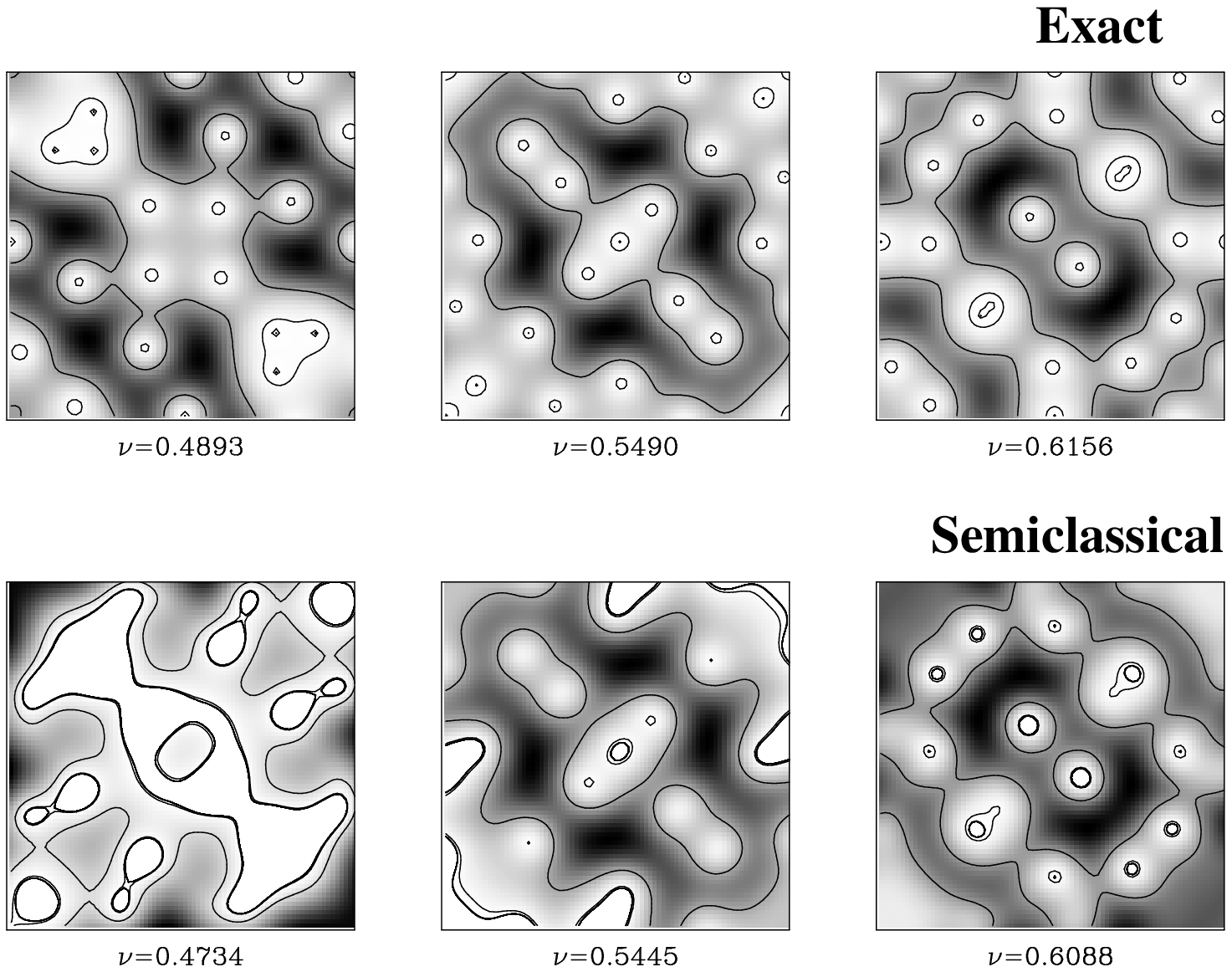}} 
\vspace{0.2cm} 
\caption{\label{gbaksem} Exact and semiclassical eigenstates in correspondence
with the zeroes in Fig.5. The gray scale and the contours are as in Fig.3. }
\end{figure} 


\begin{thebibliography}{99} 
 
\bibitem{berry1} M. V. Berry, Les Houches Lecture Notes, Summer School 
on Chaos and 
quantum Physics, M.-J. Giannoni, A. Voros, and J. Zinn-Austin, eds., 
Elsevier Science 
Publishers B. V. (1991); {\sl Proc. Roy. Soc.} {\bf A 243}, 219 (1989) 
\bibitem{voros} P. Leboeuf and A. Voros, {\sl J. Phys.} {\bf A 23}, 1765 
(1990) 
\bibitem{heller} E. J. Heller, {\sl Phys. Rev. Lett.} {\bf 53}, 1515 
(1984); Wavepacket 
Dynamics and Quantum Chaology in {\sl Chaos and Quantum Physics}, M.-J. 
Giannoni, A. 
Voros, and J. Zinn-Austin, eds., Elsevier Science Publishers, Amsterdam 
(1990) 
\bibitem{bogo1} E. B. Bogomolny {\sl Physica} {\bf D 31}, 169 (1988) 
\bibitem{vergini} D. A. Wisniacki and E. Vergini, {\sl Phys. Rev. E} 
{\bf 62}, 4513 
(2000) 
\bibitem{nonnenmacher}S. Nonnenmacher, A. Voros {\sl J. Stat. Phys.}
{\bf 92}, 431 (1998)
\bibitem{mirlin} A. D. Mirlin, Lecture course given at the International 
School {\sl Enrico Fermi} on New Directions in Quantum Chaos, Varenna, 
July 1999 (cond-mat/0006421) 
\bibitem{gutzw} M. C. Gutzwiller, {\sl J. Math. Phys.} {\bf 8}, 1979 
(1967) 
\bibitem{bogo-keat} E. B. Bogomolny and J. P. Keating, {\sl Phys. Rev. 
Lett.} {\bf 77}, 
6091 (1994) 
\bibitem{agamfish} O. Agam and S. Fishman, {\sl Phys. Rev. Lett.} {\bf 
73}, 806 (1994) 
\bibitem{sara-simo} M. Saraceno and F. Simonotti, {\sl Phys. Rev. E} 
{\bf 61}, 6527 
(2000) 
\bibitem{pseudo} M. V. Berry and J. P. Keating, {\sl J. Phys.} {\bf A 
23}, 4839 (1990) 
\bibitem{bogo2} E. B. Bogomolny, {\sl Nonlinearity} {\bf 5}, 805 (1992) 
\bibitem{bbl} E. B. Bogomolny, O. Bohigas and P. Leboeuf, {\sl Phys. Rev. 
Lett.} {\bf 68}, 2726 (1992) 
\bibitem{marsden} M. Marsden, {\sl Geometry of Polynomials}, American 
Mathematical 
Society, Providence, Rhode Island (1966). 
\bibitem{berry-keat} M. V. Berry and J. P. Keating, {\sl Proc. Roy. 
Soc.} {\bf A 437}, 
151 (1992) 
\bibitem{ozorio}A. M. Ozorio de Almeida, {\sl Phys. Rep. }{\bf 295},265 (1998)
\bibitem{fishman} S. Fishman, B. Georgeot and R. E. Prange, {\sl J. 
Phys.} {\bf A 29}, 
919 (1996); S. Fishman in {\sl Supersymmetry and Trace Formulae: Chaos 
and Disorder}, 
Lerner {\sf et al.} Eds., NATO ASI {\bf B} Vol. 370, Kluwer Academic, 
New York (1999) 
\bibitem{hannay} J. H. Hannay and M. V. Berry, {\sl Physica} {\bf 1D}, 
267 (1980) 
\bibitem{baker} M. Saraceno and A. Voros, {\sl Physica D} {\bf 79}, 206 
(1994) 
 
\end{thebibliography}
\end{document}